\documentclass[a4paper]{article}
\usepackage{RR}
\usepackage{hyperref}

\RRdate{Novembre 2008}

\RRauthor{
Pierre Genev\`es\thanks{CNRS}
   \and
Nabil Laya\"ida
\and
Vincent Quint
}

\authorhead{Genev\`es, Laya\"ida, \& Quint}

\RRtitle{Ensuring Query Compatibility with Evolving XML Schemas}
\RRetitle{Ensuring Query Compatibility with Evolving XML Schemas}
\titlehead{Ensuring Query Compatibility with Evolving XML Schemas}

\RRresume{Durant le cycle de vie d'une application XML, \`a la fois les sch\'emas et les requ\^etes sont amen\'es \`a \'evoluer d'une version \`a une autre. Les \'evolutions de sch\'emas peuvent affecter les r\'esultats des requ\^etes et potentiellement la validit\'e des donn\'ees produites. De nos jours, un vrai d\'efi consiste \`a \'evaluer et \`a prendre en compte l'impact de ces changements dans des applications XML qui \'evoluent rapidement.

Cet article propose un cadre logique et un outil pour la v\'erification des compatibilit\'es ascendante et descendante des sch\'emas et des requ\^etes. Tout d'abord, il permet d'analyser les relations entre les sch\'emas. Ensuite, il permet au concepteur XML d'identifier les requ\^etes qui doivent \^etre reformul\'ees afin de produire les r\'esultats attendus \`a travers les versions successives  des sch\'emas. Enfin, il permet d'examiner de mani\`ere plus pr\'ecise l'impact des changements des sch\'emas sur les requ\^etes, facitilitant de ce fait leur formulation. 
  
}
\RRabstract{During the life cycle of an XML application, both schemas and queries may change from one version to another. Schema evolutions may affect query results and potentially the validity of produced data. Nowadays,
a challenge is  to assess and accommodate the impact of theses changes in rapidly evolving XML applications. 

This article proposes a logical framework and tool  for verifying forward/backward compatibility issues involving schemas and queries. First, it allows analyzing relations between schemas. Second, it allows XML designers to identify queries that must be reformulated  in order to produce the expected results
across successive schema versions. Third, it allows examining more precisely 
the impact of schema changes over queries, therefore facilitating their reformulation.
}

\RRmotcle{XML, Schema, Requ\^etes, XPath, Evolution, Compatibilit\'e, Analyse}
\RRkeyword{XML, Schema, Queries, XPath, Evolution, Compatibility, Analysis}

\RRprojets{Wam}

\RRtheme{\THSym} 
\RCGrenoble

\usepackage{amsmath,amssymb}
\usepackage{stmaryrd} 
\usepackage{tikz} 
  
\usepackage{pgf} 
\usepackage{multirow}
\usepackage{fancyvrb}

\renewcommand{\phi}{\varphi}

\newcommand{\step}[2]{\text{{#1}::}{#2}}
\newcommand{\axis}[1]{\text{#1}}
\newcommand{\axisvar}{\emph{axis}}

\newcommand{\pinter}{\cap}
\newcommand{\qualif}[2]{{#1}\text{[}{#2}\text{]}}
\newcommand{\op}[1]{\mathbin{\text{\small {#1}}}}

\newcommand{\et}{\wedge}

\newcommand{\atomprop}{p}

\newcommand{\typebind}[2]{\texttt{let } \overline{{#1}.{#2}} \texttt{ in }}

\newcommand{\singletypebind}[2]{{\texttt{let } {#1}.{#2}} \texttt{ in }}

\newcommand{\tou}{\mid}

\newcommand{\nullable}[1]{\text{\emph{nullable}}({#1})}

\newcommand{\Tsucc}[2]{\text{\emph{s}}_{#2}({#1})}

\newcommand{\rewritesinto}{\rightsquigarrow}

\newcommand{\eqdef}{\stackrel{\text{\tiny def}}{=}}

\newcommand{\syntaxdef}{\mathrel{::=}}

\newcommand{\syntaxtable}[1]{
  \def\entry##1[##2]##3[##4]{
    {##1} & \syntaxdef & \hspace{3cm} & \!\!\!\! \mbox{##2}
    \\    &     & {##3} & \mbox{##4} }
  \def\singleentry##1[]##2[##3]{
  {##1} & \syntaxdef & {##2} & \!\!\!\! \mbox{##3} }
  \def\oris##1[##2]{
    \\    & 
       & {##1} & \mbox{##2} }
  \def\orisopt##1[##2]{
    \\ \left(   & 
       & {##1} & \mbox{##2} \right) }
  \begin{array}{rcll}
  #1
  \end{array}
  }
\newcommand{\smallsyntax}[1]{\[\syntaxtable{#1}\]}

\newcommand{\treetypevar}{\tau} 

\newcommand{\ie}{{\text{{\emph{i.e.}}}} }
\newcommand{\eg}{{\text{{\emph{e.g.}}}} }

\newcommand{\nodetestvar}{\text{\emph{nt}}}
\newcommand{\locpathvar}{\emph{path}}

\newcommand{\qualifvar}{\emph{qualifier}}
\newcommand{\exprvar}{\emph{query}}

\newcommand{\xpathfun}[2]{\text{#1}({#2})}

\newcommand{\treetypelabelwithattributes}[3]{{#1}\texttt{(}{#2}\texttt{)}\texttt{[}{#3}\texttt{]}}
\newcommand{\emptyseq}{\texttt{()}}

\newcommand{\attributeexpr}{\text{\emph{a}}}
\newcommand{\attributeconj}{\text{\emph{list}}}

\newcommand{\attributevar}[1]{{#1}}
\newcommand{\optionalattributevar}[1]{{#1}\texttt{?}}
\newcommand{\attributeexprvar}{\text{\emph{a}}}
\newcommand{\negateotheratts}[1]{\text{notothers(}{#1})}

\newcommand{\nodelabelvar}{l}

\newcommand{\translatetype}[3]{\text{tr}(#1)_{#2}^{#3}}
\newcommand{\attrtomu}[1]{\text{tra}( #1)}

\newcommand{\programvar}{p}

\newcommand{\ctrue}{\texttt{T}}
\newcommand{\cfalse}{\texttt{F}}

\newcommand{\cimplies}{\text{ }\texttt{=>}\text{ }}
\newcommand{\cequiv}{\text{ }\texttt{<=>}\text{ }}

\newcommand{\cneg}{\tilde{}}

\newcommand{\corop}{\text{ }\texttt{|}\text{ }}
\newcommand{\candop}{\text{ }\texttt{\&}\text{ }}
\newcommand{\ccontextsymb}{\texttt{\#}}

\newcommand{\cprog}[1]{\texttt{#1}}

\newcommand{\cvar}{\texttt{\$X}}
\newcommand{\cvary}{\texttt{\$Y}}

\newcommand{\clet}{\texttt{let}}
\newcommand{\cin}{\texttt{in}}

\newcommand{\cemod}[1]{\texttt{<}#1\texttt{>}}

\newcommand{\cpredicatevar}{\text{\emph{predicate}}}

\newcommand{\oneormoresep}[1]{\langle #1 \rangle^\varoplus}

\newcommand{\cselect}[1]{\texttt{select}(\texttt{"}#1\texttt{"})}
\newcommand{\cselectcontext}[2]{\texttt{select}(\texttt{"}#1\texttt{"}, #2)}
\newcommand{\cexists}[1]{\texttt{exists}(\texttt{"}#1\texttt{"})}
\newcommand{\cexistscontext}[2]{\texttt{exists}(\texttt{"}#1\texttt{"}, #2)}

\newcommand{\ctype}[2]{\texttt{type}(\texttt{"}#1\texttt{"}, #2)}
\newcommand{\ctypetag}[4]{\texttt{type}(\texttt{"}#1\texttt{"}, #2, #3, #4)}

\newcommand{\filenamevar}{\text{\emph{f}}}

\newcommand{\cnonempty}[2]{\texttt{non\_empty}(\texttt{"}#1\texttt{"}, #2)}

\newcommand{\celem}[1]{\texttt{element}(#1)}
\newcommand{\cattr}[1]{\texttt{attribute}(#1)}
\newcommand{\caddedelement}[2]{\texttt{added\_element}(#1,#2)}
\newcommand{\caddedattribute}[2]{\texttt{added\_attribute}(#1,#2)}
\newcommand{\cexclude}[1]{\texttt{exclude}(#1)}
\newcommand{\cforwardincompatible}[2]{\texttt{forward\_incompatible}(#1,#2)}
\newcommand{\cbackwardincompatible}[2]{\texttt{backward\_incompatible}(#1,#2)}

\newcommand{\cnewelementnames}[3]{\texttt{new\_element\_name}(\texttt{"}#1\texttt{"},#2,#3)}
\newcommand{\cnewregions}[3]{\texttt{new\_region}(\texttt{"}#1\texttt{"},#2,#3)}

\newcommand{\cnewcontents}[3]{\texttt{new\_content}(\texttt{"}#1\texttt{"},#2,#3)}

\newcommand{\ccnewelementnames}[4]{\texttt{new\_element\_name}(\texttt{"}#1\texttt{"},\texttt{"}#2\texttt{"},\texttt{"}#3\texttt{"}, #4)}
\newcommand{\ccnewregions}[4]{\texttt{new\_region}(\texttt{"}#1\texttt{"},\texttt{"}#2\texttt{"},\texttt{"}#3\texttt{"}, #4)}

\newcommand{\ccnewcontents}[4]{\texttt{new\_content}(\texttt{"}#1\texttt{"},\texttt{"}#2\texttt{"},\texttt{"}#3\texttt{"}, #4)}

\newcommand{\cdescendant}[1]{\texttt{descendant}(#1)}
\newcommand{\cancestor}[1]{\texttt{ancestor}(#1)}
\newcommand{\cdescendantorself}[1]{\texttt{descendant-or-self}(#1)}
\newcommand{\cancestororself}[1]{\texttt{ancestor-or-self}(#1)}
\newcommand{\cfollowing}[1]{\texttt{following}(#1)}
\newcommand{\cpreceding}[1]{\texttt{preceding}(#1)}

\newcommand{\call}{\texttt{\_all}}
\newcommand{\coldcomplement}{\texttt{\_old\_complement}}

\newcommand{\custompredicatevar}{\text{\emph{predicate-name}}}

\newcommand{\multistep}[2]{(#1)^{#2}}

\newcommand{\logicalspec}{\text{\emph{spec}}}
\newcommand{\predicatedefinitions}{\text{\emph{def}}}

\begin{document}   
\RRNo{6711}
\makeRR 

\section{Introduction}

XML is now commonplace on the web and in many information systems where it is
used for representing all kinds of information resources, ranging from simple
text documents such as RSS or Atom feeds to highly structured databases.
In these dynamic environments, not only data are changing steadily but their
schemas also get modified to cope with the evolution of the real world entities
they describe.

Schema changes raise the issue of data consistency. Existing documents and
data that were valid with a certain version of a schema may become invalid
on a new version of the schema (forward incompatibility). Conversely, new
documents created with the latest version of a schema may be invalid on some
previous versions (backward incompatibility).

In addition, schemas may be written in different languages, such as DTD,
XML Schema, or Relax-NG, to name only the most popular ones. And it is common
practice to describe the same structure, or new versions of a structure,
in different schema languages. Document formats developed by W3C provide a
variety of examples: XHTML 1.0 has both DTDs and XML Schemas, while XHTML 2.0
has a Relax-NG definition; the schema for SVG Tiny 1.1 is a DTD, while
version 1.2 is written in Relax-NG; MathML 1.01 has a DTD, MathML 2.0 has
both a DTD and an XML Schema, and MathML 3.0 is developed with a Relax-NG
schema and is expected to have also a DTD and an XML Schema.
An issue then is to make sure that schemas written in different languages are
equivalent, \ie they describe the same structure, possibly with some differences
due to the expressivity of the language \cite{murata-toit05}.
Another issue is to clearly identify the differences between two
versions of the same schema expressed in different languages. Moreover, the
issues of forward and backward compatibility of instances obviously remain
when schema languages change from a version to another.

Validation, and then compatibility, is not the only purpose of a schema.
Validation is usually the first step for safe processing of documents and
data. It makes sure that documents and data are structured as expected and
can then be processed safely. The next step is to actually access and select
the various parts to be handled in each phase of an application. For this,
query languages play a key role. As an example, when transforming a document
with XSL, XPath queries are paramount to locate in the original document
the data to be produced in the transformed document.

Queries are affected by schema evolutions. The structures they return may
change depending on the version of the schema used by a document. When
changing schema, a query may return nothing, or something different from
what was expected, and obviously further processing based on this query is
at risk.

These observations highlight the need for evaluating precisely
and safely the impact of schema evolutions on existing and future instances
of documents and data. They also show that it is important for software
engineers to precisely know what parts of a processing chain have to be
updated when schemas change. In this paper we focus on the XPath query
language which is used in many situations while processing XML documents
and data. The XSL transformation language was already mentioned, but XPath
is also present in XLink and XQuery for instance.

%

\subsection*{Related Work}

Schema evolution is an important topic and has been extensively explored in the context of relational, object-oriented, and XML databases. Most of the previous work for XML query reformulation is approached through reductions to relational problems \cite{beyer-sigmod05}. 
This is because schema evolution was considered as a storage problem where the priority consists in ensuring data consistency across multiple relational schema versions. In such settings, two distinct schemas and an explicit description of the mapping between them are assumed as input. The problem then consists in reformulating a query expressed in terms of one schema into a semantically equivalent query in terms of the other schema: see \cite{deutsch-tcs05,yu-vldb05} and more recently \cite{deutsch-vldb08} with references thereof.

In addition to the fundamental differences between XML and the relational data model, in the more general case of XML processing, schemas constantly evolve in a distributed, independent, and unpredictable environment. The relations between different schemas are not only unknown but hard to track. 
In this context, one priority  is to help maintaining query consistency during these evolutions, which is still considered as a challenging problem \cite{sedlar-sigmod05}. 


The work found in \cite{moro-www07}  discusses the impact of evolving XML schemas on query reformulation.
Based on a taxonomy of XML schema changes during their evolution, the authors provide informal -- not exact nor systematic -- guidelines for writing queries which are less sensitive to schema evolution. In fact, studying query reformulation requires at least the ability to analyze the relationship between queries. 
For this reason,  a closely related work is the problem of determining query containment and satisfiability under type constraints \cite{benedikt-pods05,geneves-pldi07}. The work found in \cite{benedikt-pods05} studies the complexity of XPath emptiness and containment for various fragments (see \cite{benedikt-unpublished08} and references thereof for a survey). 

The main distinctive idea pursued in this paper is to develop a logical approach for guiding schema and query evolution. In contrast to the classical use of logics for proving properties such as query emptiness or equivalence \cite{benedikt-pods05,geneves-pldi07}, the goal here is different in that we seek to provide the necessary tools to produce relevant knowledge when such relations do not hold. 





\subsection*{Outline}
The rest of this paper is organized as follows: the next section introduces our framework, Section~\ref{sec:logical-formulation} presents its underlying logic, and Section~\ref{sec:predicates} presents predicates for characterizing the impact of schema changes. We report on experiments on realistic scenarios in Section~\ref{sec:applications-xml-evolution} before we conclude in Section~\ref{conclusion}.

\section{Analysis Framework} \label{sec:framework}
Our framework allows the automatic verification of properties related to XML schema and query evolution.
In particular, it offers the possibility of checking fine-grained properties on the behavior of queries with respect to successive versions of a given schema. 
The system can be used for checking whether schema evolutions require a particular query to be updated. Whenever schema evolutions may induce query malfunctions, the system is able to generate annotated XML documents that exemplify bugs, with the goal of  helping the programmer to understand and properly overcome undesired effects of schema evolutions.
 
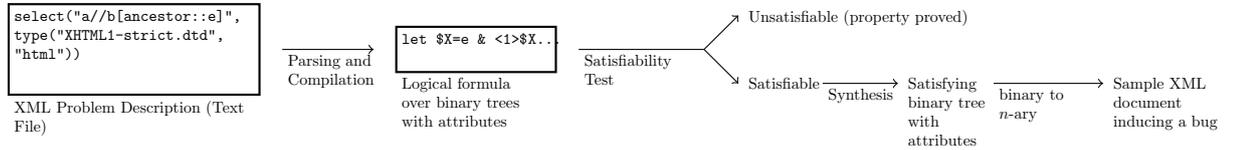
\begin{figure*}
\begin{center}
\begin{tikzpicture}[scale=0.6] at (0,0)
\newcommand{\opaq}{0.8}

\begin{scope} [xshift=0cm, yshift=0cm]
\draw [thick, black] (0,-2) -- (5.5,-2) -- (5.5,0) -- (0,0) -- cycle;
\draw (0,0) node [anchor=north west,  text width=5cm, text badly ragged, at start] 
{ \texttt{select("a//b[ancestor::e]", type("XHTML1-strict.dtd", "html"))} \\
}; 
	\begin{scope} [xshift=0cm, yshift=-2cm]
	\draw (0,0) node [anchor=north west,  text width=5.5cm, text badly ragged, at start] {XML Problem Description (Text File)}; 
	\end{scope}
\end{scope}

\begin{scope} [xshift=6cm, yshift=-1cm]
\draw[->] (0,0) -- (2,0);
\draw (0,0) node [anchor=north west,  text width=5cm, text badly ragged, at start] {Parsing and\\Compilation}; 
\end{scope}

\begin{scope} [xshift=8.5cm, yshift=-0.5cm]
\draw [thick, black] (0,-1) -- (3.5,-1) -- (3.5,0) -- (0,0) -- cycle;
\draw (0,0) node [anchor=north west,  text width=5cm, text badly ragged, at start] 
{ \texttt{let \$X=e \& <1>\$X...} \\
}; 
	\begin{scope} [xshift=0cm, yshift=-1cm]
	\draw (0,0) node [anchor=north west,  text width=3cm, text badly ragged, at start] {Logical formula over binary trees with attributes}; 
	\end{scope}
\end{scope}

\begin{scope} [xshift=12.5cm, yshift=-1cm]
\draw[-] (0,0) -- (2.75,0);
\draw (0,0) node [anchor=north west,  text width=2.5cm, text badly ragged, at start] {Satisfiability Test}; 

\draw[->] (2.75,0) -- (3.5,0.75);
	\begin{scope} [xshift=3.6cm, yshift=1cm]
	\draw (0,0) node [anchor=north west,  text width=5cm, text badly ragged, at start] {Unsatisfiable (property proved)}; 
	\end{scope}

\draw[->] (2.75,0) -- (3.5,-0.75);
	\begin{scope} [xshift=3.6cm, yshift=-0.5cm]
	\draw (0,0) node [anchor=north west,  text width=3cm, text badly ragged, at start] {Satisfiable};

		\begin{scope} [xshift=1.75cm, yshift=-.25cm]	
		\draw[->] (0,0) -- (1.65,0);
			\begin{scope} [xshift=0cm, yshift=0cm]	
			\draw (0,0) node [anchor=north west,  text width=2cm, text badly ragged, at start]{Synthesis}; 
			\end{scope}

			\begin{scope} [xshift=1.75cm, yshift=0.25cm]	
			\draw (0,0) node [anchor=north west,  text width=2.25cm, text badly ragged, at start]{Satisfying binary tree with attributes}; 
			\end{scope}

	\begin{scope} [xshift=3.75cm, yshift=0cm]	
		\draw[->] (0,0) -- (2.4,0);
					\begin{scope} [xshift=0cm, yshift=0cm]	
			\draw (0,0) node [anchor=north west,  text width=2.25cm, text badly ragged, at start]{binary to $n$-ary}; 
			\end{scope}

			\begin{scope} [xshift=2.5cm, yshift=0.25cm]	
			\draw (0,0) node [anchor=north west,  text width=2.5cm, text badly ragged, at start]{Sample XML document inducing a bug}; 
			\end{scope}
		\end{scope}
		\end{scope}
	\end{scope}

\end{scope}
\end{tikzpicture}
\end{center}
\caption{Framework Overview.}\label{fig:framework}.
\end{figure*}

For these purposes, our framework relies on the combination and joint use of several contributions:
\begin{itemize}
\item an extension of the logic introduced in  \cite{geneves-pldi07} to deal with XML attributes (Sections~\ref{sec:data-model} and~\ref{sec:logical-formulation});
\item a set of logical features and high-level predicates specifically designed for studying and characterizing schema and query compatibility issues when schemas evolve   (Section~\ref{sec:predicates});
\item a range of applications and procedures to cope with schema and query evolution (Section~\ref{sec:applications-xml-evolution});
\item a full implementation of the whole system, including:
	\begin{itemize}
	\item a parser for reading the problem description (text file), which in turn use specific parsers for schemas (Section~\ref{sec:tree-type-expressions}), queries (Section~\ref{sec:queries}), logical formulas (Section~\ref{sec:formulas}), and predicates (Section~\ref{sec:predicates});
	\item compilers for translating schemas and queries into their logical representations (Sections~\ref{sec:xpath-compilation} and~\ref{sec:xml-schemas-compilation});
	\item an optimized solver first described in \cite{geneves-pldi07,geneves-rr08} for checking satisfiability of logical formulas in time $2^{O(n)}$ where $n$ is the formula size;
	\item and a counter example XML tree generator (described in \cite{geneves-rr08}).
	\end{itemize}
\end{itemize}
Figure~\ref{fig:framework} illustrates how the previous software components are combined and used together, in a simplified overview of the global framework.
We next introduce the data model we consider for XML documents, schemas and queries.

 \label{sec:data-model}
\subsection{XML Trees with Attributes}
An XML document is considered as a finite tree of unbounded depth and arity, with two kinds of nodes respectively named elements and attributes.  In such a tree, an element may have any number of children elements, and may carry zero, one or more attributes. Attributes are leaves. Elements are ordered whereas attributes are not, as illustrated on Figure~\ref{fig:xml-tree}. In this paper, we focus on the nested structure of elements and attributes, and ignore XML data values.

\subsection{Type Constraints} \label{sec:tree-type-expressions}

As an internal representation for tree grammars, we consider regular tree type expressions (in the manner of \cite{hosoya-toplas05}), extended with constraints over attributes. Assuming a set of variables ranged over by $x$, we define a tree type expression as follows:
\smallsyntax{
\entry      \treetypevar     [tree type expression] 
             \emptyset              [empty set]
\oris        \emptyseq       [empty sequence] 
\oris        \treetypevar \tou \treetypevar    [disjunction]
\oris        \treetypevar, \treetypevar    [concatenation]
\oris        \treetypelabelwithattributes{\nodelabelvar}{\attributeexprvar}{\treetypevar}      [element definition]
\oris        x  	       [variable] 
\oris        \typebind{x}{\treetypevar} \treetypevar [binder]
}
We impose a usual restriction on the recursive use of variables: we allow unguarded (\ie not enclosed by a label) recursive uses of variables, but restrict them to tail positions\footnote{For instance, ``$\singletypebind{x}{\treetypelabelwithattributes{\nodelabelvar}{\attributeexprvar}{\treetypevar},x  \tou \emptyseq} x$'' is allowed.}. With that restriction, tree types expressions define regular tree languages. In addition, an element definition may involve simple attribute expressions that describe which attributes the defined element may (or may not) carry:
	\smallsyntax{
	\entry \attributeexpr    [attribute expression]
	\emptyseq [empty list] 
	\oris \attributeconj \tou \attributeexpr [disjunction]\\
	\entry \attributeconj    [attribute list]
	\attributeconj, \attributeconj [commutative concatenation]
	\oris \optionalattributevar{\nodelabelvar} [optional attribute]
	\oris \attributevar{\nodelabelvar} [required attribute]
	\oris \neg \attributevar{\nodelabelvar} [prohibited attribute]
}
Our tree type expressions capture most of the schemas in use today \cite{murata-toit05, neven-www05}.
In practice, our system provides parsers that convert DTDs, XML Schemas, and Relax NGs to this internal tree type representation. Users may thus  define constraints over XML documents with the language of their choice, and, more importantly, they may refer to most existing schemas for use with the system.


\subsection{Queries}
\label{sec:queries}

The set of XPath expressions we consider is given by the syntax shown on Figure~\ref{fig:xpath-syntax}. The semantics of XPath expressions is described in \cite{xpath}, and more formally in \cite{wadler}. We observed that, in practice, many XPath expressions contain syntactic sugars that can also fit into this fragment.
Figure~\ref{fig:xpath-sugars} presents how our XPath parser rewrites some commonly found XPath patterns into the fragment of Figure~\ref{fig:xpath-syntax}, where the notation $\multistep{ \step{\axisvar}{\nodetestvar}}{k}$ stands for the composition of $k$ successive path steps of the same form:
$\underbrace{ \step{\axisvar}{\nodetestvar}/.../ \step{\axisvar}{\nodetestvar}}_{k ~ \text{steps}}$.

\begin{figure}[h]
\smallsyntax{
\entry   \exprvar     []
             /\locpathvar              [absolute path]
\oris        \locpathvar              [relative path]
\oris        \exprvar \mid \exprvar [union] 
\oris        \exprvar \pinter \exprvar [intersection] \\
\entry      \locpathvar          []
             \locpathvar/\locpathvar      [path composition]
\oris        \qualif{\locpathvar}{\qualifvar} [qualified path]
\oris        \step{\axisvar}{\nodetestvar} [step] \\
\entry       \qualifvar         []
    					\qualifvar \op{and} \qualifvar [conjunction]
\oris					\qualifvar \op{or} \qualifvar  [disjunction]
\oris					\xpathfun{not}{\qualifvar}   [negation]
\oris					\locpathvar            [path] 
\oris					\locpathvar/@{\nodetestvar}             [attribute path] 
\oris					@{\nodetestvar}           [attribute step] 
\\
\entry      \nodetestvar          [node test]
            \sigma      [node label]
\oris        * [any node label] \\
\entry    \axisvar [tree navigation axis]
             \axis{self}  
     \mid \axis{child} 
     \mid  \axis{parent}[]
\oris       \axis{descendant} 
\mid        \axis{ancestor} []
\oris  \axis{descendant-or-self} []
\oris  \axis{ancestor-or-self}  []
\oris       \axis{following-sibling} []
\oris  \axis{preceding-sibling}  []
\oris       \axis{following} 
\mid \axis{preceding} [] \\
}
\caption{XPath Expressions.}\label{fig:xpath-syntax}
\end{figure}

\begin{figure}[h]
\begin{align*}
\qualif{\nodetestvar}{\xpathfun{position}{}=1} ~&~ \rewritesinto \qualif{\nodetestvar}{\xpathfun{not}{\step{\axis{preceding-sibling}}{\nodetestvar}}} \\
\qualif{\nodetestvar}{\xpathfun{position}{}=\xpathfun{last}{}} ~&~ \rewritesinto \qualif{\nodetestvar}{\xpathfun{not}{\step{\axis{following-sibling}}{\nodetestvar}}} \\
\qualif{\nodetestvar}{\xpathfun{position}{}=\underbrace{k}_{k>1}}  ~&~ \rewritesinto  \qualif{\nodetestvar}{ \multistep{\step{\axis{preceding-sibling}}{\nodetestvar}}{k-1}} \\
\xpathfun{count}{\locpathvar}=0 ~&~ \rewritesinto \xpathfun{not}{\locpathvar} \\
\xpathfun{count}{\locpathvar}>0 ~&~ \rewritesinto \locpathvar\\
\xpathfun{count}{\nodetestvar}>\underbrace{k}_{k>0} ~&~ \rewritesinto \nodetestvar/\multistep{\step{\axis{following-sibling}}{\nodetestvar}}{k}
\end{align*}
\begin{multline*}
\qualif{\step{\axis{preceding-sibling}}{*}}{\xpathfun{position}{}=\xpathfun{last}{} \op{and} \qualifvar } \\ \rewritesinto \qualif{\step{\axis{preceding-sibling}}{*}}{\xpathfun{not}{\step{\axis{preceding-sibling}}{*}} \op{and} \qualifvar } 
\end{multline*}
\caption{Syntactic Sugars and their Rewritings.}\label{fig:xpath-sugars}
\end{figure}

\section{Logical Setting} \label{sec:logical-formulation}
\subsection{Logical Data Model}  

It is well-known that there exist bijective encodings between unranked trees (trees of unbounded arity) and binary trees. Owing to these encodings binary trees may be used instead of unranked trees without loss of generality.  In the sequel, we rely on a simple ``first-child \& next-sibling'' encoding of unranked trees. In this encoding, the first child of an element node is preserved in the binary tree representation, whereas siblings of this node are appended as right successors in the binary representation. Attributes are left unchanged by this encoding.  For instance, Figure~\ref{fig:binary-tree} presents how the sample tree of Figure~\ref{fig:xml-tree} is mapped.
\definecolor{mydarkblue}{rgb}{0,0.08,0.45} 
\newcommand{\changingcolor}{gray}
\newcommand{\metiquette}[1]{\begin{tiny}{#1}\end{tiny}}
\newcommand{\setiquette}[1]{\begin{small}{#1}\end{small}}
\newcommand{\labelstyle}[1]{\texttt{#1}}
\newcommand{\attdefvalue}{\textvisiblespace}
\newcommand{\attcolor}{mydarkblue}
\begin{figure}[h]
\begin{center}
\begin{tikzpicture}[scale=1.1] at (0,0)

\begin{scope}[xshift=-3cm, yshift=-0.9cm, scale=0.6]
 \draw [black, font=\scriptsize, text badly ragged, at start, text width=5cm]  (16,4) node {\texttt{<r \textcolor{\attcolor}{c="\attdefvalue" a="\attdefvalue" b="\attdefvalue"}>} \\~~\texttt{<s \textcolor{\attcolor}{d="\attdefvalue"}>} \\~~~~\texttt{<v/><w/><x \textcolor{\attcolor}{e="\attdefvalue"}/>} \\~~\texttt{</s>} \\~~\texttt{<t/>}\\~~\texttt{<u/>} \\\texttt{</r>}};
 \draw [black] (12.1,5.5) -- (16.5,5.5) -- (16.5,2.5) -- (12.1,2.5) -- cycle;
 \draw[black, font=\scriptsize, text badly ragged, at start, text width=5cm] (16.25, 2.25) node (not) {\textbf{XML Notation}};
\end{scope}
		
\begin{scope}[xshift=0cm, yshift=0cm, scale=0.6]
	\draw [\attcolor, dashed]  (14.9,3.6) .. controls +(40:-0.5cm) and +(40:0.1cm) .. (14,3.9); 
	\draw (13.8, 4.1) node (att3) [text= \attcolor] {\labelstyle{a}};

	\draw [\attcolor, dashed]  (15.5,3.6) .. controls +(40:1.5cm) and +(-40:-0.1cm) .. (17,3.9); 
	\draw (17.2, 3.7) node (att1) [text= \attcolor] {\labelstyle{b}};
	
	\draw [\attcolor, dashed]  (15.5,3.6) .. controls +(60:2.5cm) and +(4:-0.3cm) .. (18,3.7); 
	\draw (18.2, 3.5) node (att2) [text= \attcolor] {\labelstyle{c}};
	
	\draw [\attcolor, dashed]  (13.5,2.5) .. controls +(40:-0.5cm) and +(40:0.1cm) .. (12.5,3); 
	\draw (12.5, 3.2) node (att4) [text= \attcolor] {\labelstyle{d}};

	\draw [\attcolor, dashed]  (15,1.2) .. controls +(40:-0.5cm) and +(-5:0.1cm) .. (16.4,0.7); 
	\draw (16.6, 0.7) node (att4) [text= \attcolor] {\labelstyle{e}};

	\draw (15.25, 3.5) node (r) [text=black] {\labelstyle{r}};

	\draw (13.75, 2.5) node (s1) [text=black] {\labelstyle{s}};
	\draw (15.25, 2.5) node (s2) [text=black] {\labelstyle{t}};		
	\draw (16.75, 2.5) node (s3) [text=black] {\labelstyle{u}};  

	\draw (12.5, 1.5) node (f1) [text=black] {\labelstyle{v}};
	\draw (13.75, 1.5) node (f2) [text=black] {\labelstyle{w}};
	\draw (15, 1.5) node (f3) [text=black] {\labelstyle{x}};
	
	\draw (r) -- (s3);
	\draw (r) -- (s1);
	\draw (r) -- (s2);
	\draw (s1) -- (f3);
	\draw (s1) -- (f2);
	\draw (s1) -- (f1);

\end{scope}
\end{tikzpicture}
\end{center}
\caption{Sample XML Tree with Attributes.}\label{fig:xml-tree}
\end{figure}
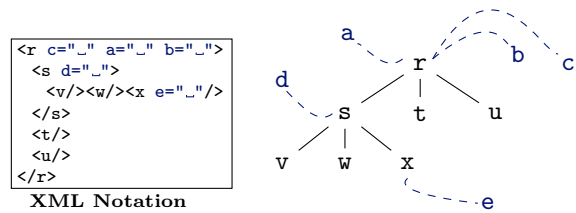
\begin{figure}[h]
\begin{center}
\begin{tikzpicture}[scale=1.1] at (0,0)

\begin{scope}[xshift=0cm, yshift=-4.5cm, scale=0.6]
	
\begin{scope}[xshift=-0.25cm, yshift=0cm]	

	\draw [\attcolor, dashed]  (14.9,3.6) .. controls +(40:-0.5cm) and +(40:0.1cm) .. (14,3.9); 
	\draw (13.8, 4.1) node (att3) [text= \attcolor] {\labelstyle{a}};

	\draw [\attcolor, dashed]  (15.5,3.6) .. controls +(40:1.5cm) and +(-40:-0.1cm) .. (17,3.9); 
	\draw (17.2, 3.7) node (att1) [text= \attcolor] {\labelstyle{b}};
	
	\draw [\attcolor, dashed]  (15.5,3.6) .. controls +(60:2.5cm) and +(4:-0.3cm) .. (18,3.7); 
	\draw (18.2, 3.5) node (att2) [text= \attcolor] {\labelstyle{c}};
	
\end{scope}

	\draw [\attcolor, dashed]  (13.5,2.5) .. controls +(40:-0.5cm) and +(40:0.1cm) .. (12.5,3); 
	\draw (12.5, 3.2) node (att4) [text= \attcolor] {\labelstyle{d}};

\begin{scope}[xshift=1.3cm, yshift=-3.2cm]	
	\draw [\attcolor, dashed]  (13.5,2.5) .. controls +(40:-0.5cm) and +(40:0.1cm) .. (11.5,3); 
	\draw (11.45, 3.25) node (att4) [text= \attcolor] {\labelstyle{e}};
\end{scope}

	\draw (15, 3.5) node (r) [text=black] {\labelstyle{r}};

	\draw (14, 2.5) node (s1) [text=black] {\labelstyle{s}};
	\draw (15, 1.5) node (s2) [text=black] {\labelstyle{t}};		
	\draw (16, 0.5) node (s3) [text=black] {\labelstyle{u}};

	\draw (13, 1.5) node (f1) [text=black] {\labelstyle{v}};
	\draw (14, 0.5) node (f2) [text=black] {\labelstyle{w}};
	\draw (15, -0.5) node (f3) [text=black] {\labelstyle{x}};
	
	\draw (r) -- (s1);
	\draw (s1) -- (s2);
	\draw (s2) -- (s3);
	\draw (s1) -- (f1);
	\draw (f1) -- (f2);
	\draw (f2) -- (f3);	 
	
\end{scope}
\end{tikzpicture}
	\end{center}
\caption{Binary Encoding of Tree of Figure~\ref{fig:xml-tree}.}\label{fig:binary-tree}
\end{figure}
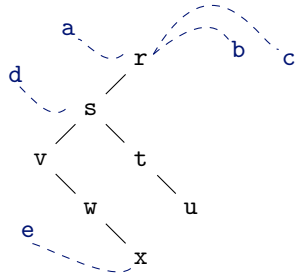

The logic we introduce below, used as the core of our framework, operates on such binary trees with attributes.

\subsection{Logical Formulas}   \label{sec:formulas}

The concrete syntax of logical formulas is shown on Figure~\ref{fig:formulas-syntax}, where the meta-syntax $\oneormoresep{X}$ means one or more occurences of $X$ separated by commas. The reader can directly use this syntax for encoding formulas as text files to be used with the system described in Section~\ref{sec:framework} \cite{implementation}. This concrete syntax is used as a single unifying notation throughout all the paper.

\begin{figure}[h]
\begin{center}
\smallsyntax{
\entry      \phi       [formula]
   	      \ctrue [true]	
\oris	      \cfalse [false]
\oris        \nodelabelvar [element name]
\oris         \atomprop [atomic proposition ]
\oris        \ccontextsymb [start context] 
\oris        \phi \corop \phi [disjunction]
\oris        \phi \candop \phi [conjunction]
\oris        \phi \cimplies \phi [implication]
\oris        \phi \cequiv \phi [equivalence]
\oris       \texttt{(} \phi \texttt{)} [parenthesized formula]
\oris        \cneg \phi [negation]
\oris        \cemod{\programvar}\phi  [existential modality]
\oris        \cemod{\nodelabelvar}\ctrue  [attribute named $\nodelabelvar$]
\oris        \cvar [variable]
\oris        \clet ~ \oneormoresep{\cvar= \phi} ~ \cin~ \phi [binder for recursion]
\oris 	      \cpredicatevar [predicate (See Section~\ref{sec:predicates})] \\
\entry \programvar [program inside modalities]
\cprog{1} [first child]
\oris \cprog{2} [next sibling]
\oris \cprog{-1} [parent]
\oris \cprog{-2} [previous sibling]
}
\end{center}
\caption{Syntax of Logical Formulas.}\label{fig:formulas-syntax}
\end{figure}

The semantics of logical formulas corresponds to the classical semantics of a $\mu$-calculus interpreted over finite tree structures. A formula is satisfiable iff there exists a finite binary tree with attributes for which the formula holds at some node. This is formally defined in \cite{geneves-pldi07}, and we review it informally below through a series of examples.

There is a difference between an element name and an atomic proposition\footnote{In practice, an atomic proposition must start with a ``\_''.}: an element has one and only one element name, whereas it can satisfy multiple atomic propositions. We use atomic propositions to attach specific information to tree nodes, not related to their XML labeling.
For example, the start context (a reserved atomic proposition) is used to mark the starting context nodes for evaluating XPath expressions.

The logic uses programs for navigating in binary trees: the program \cprog{1} allows to navigate from a node down to its first successor and the program \cprog{2} for navigating from a node down to its second successor. The logic also features converse programs \cprog{-1} and \cprog{-2} for navigating upward in binary trees, respectively from the first successor to its parent and from the second successor to its previous sibling. 
Table~\ref{table:sample-formulas} gives some simple formulas using modalities for navigating in binary trees, together with sample satisfying trees, in binary and unranked tree representations.

\begin{table}
\begin{center}
\begin{tabular}{|c|c|c|}
\hline
\textbf{Sample Formula} & \textbf{Tree} & \textbf{XML} \\ 
\hline 
\texttt{a \& <1>b}
&
\begin{tikzpicture}[scale=0.75]
\draw (0,1) node (a) {\textbf{a}};
\draw (-1,0) node (b) {b};
\draw [black] (a) -- (b);
\end{tikzpicture}
&
\texttt{<a><b/></a>} \\
\hline
\texttt{a \& <1>(b \& <2>c)}
&
\begin{tikzpicture}[scale=0.75]
\draw (0,1) node (a) {\textbf{a}};
\draw (-1,0) node (b) {b};
\draw (0,-1) node (c) {c};
\draw [black] (a) -- (b) -- (c);
\end{tikzpicture}
&
\texttt{<a><b/><c/></a>} \\
\hline
\texttt{e \& <-1>(d \& <2>g)}
&
\begin{tikzpicture}[scale=0.75]
\draw (0,1) node (d) {d};
\draw (-1,0) node (e) {\textbf{e}};
\draw (1,0) node (g) {g};
\draw [black] (g) -- (d) -- (e);
\end{tikzpicture}
&
\texttt{<d><e/></d><g/>} \\
\hline

\texttt{f \& <-2>(g \& \~{}<2>T)}
&
none
&
none \\
\hline
\end{tabular}
\end{center}
\caption{Sample Formulas and Satisfying Trees.}\label{table:sample-formulas}
\end{table}


The logic allows expressing recursion in trees through the recursive binder.  For example the recursive formula: \begin{center}\texttt{let \$X = b | <2>\$X in \$X}\end{center} means that either the current node is named \texttt{b} or there is a sibling of the current node which is named \texttt{b}.  For this purpose, the variable \texttt{\$X} is bound to the subformula \texttt{b | <2>\$X} which contains an occurence of \texttt{\$X} (therefore defining the recursion). The scope of this binding is the subformula that follows the ``\texttt{in}'' symbol of the formula, that is \texttt{\$X}. The entire formula can thus be seen as a compact recursive notation for a infinitely nested formula of the form: \begin{center}\texttt{b | <2>(b | <2>(b | <2>(...)))}\end{center}

\noindent Recursion allows expressing global properties. For instance, the recursive formula: \begin{center}\texttt{\~{} let \$X = a | <1>\$X | <2>\$X in \$X} \end{center} expresses the absence of  nodes named \texttt{a} in the whole subtree of the current node (including the current node). Furthermore, the fixpoint operator makes possible to bind several variables at a time, which is specifically useful for expressing mutual recursion. For example, the mutually recursive formula: 
\begin{center}
\begin{tabular}{l}
\texttt{let }\\
~~~\texttt{\$X = (a \& <2>\$Y) | <1>\$X | <2>\$X,} \\
~~~\texttt{\$Y =  b | <2>\$Y} \\
\texttt{in \$X} 
\end{tabular}
\end{center}
asserts that there is a node somewhere in the subtree such that this node is named \texttt{a} and it has at least one sibling which is named \texttt{b}.  Binding several variables at a time provides a very expressive yet succinct notation for expressing mutually recursive structural patterns (that are common in XML Schemas, for instance).

From a theoretical perspective, the recursive binder $\clet ~ \cvar= \phi ~ \cin~ \phi$ corresponds to the fixpoint operators of the $\mu$-calculus. It is shown in \cite{geneves-pldi07} that the least fixpoint and the greatest fixpoint operators of the $\mu$-calculus coincide over finite tree structures, for a restricted class of formulas called \emph{cycle-free}  formulas. Translations of XPath expressions and schemas presented in this paper always yield cycle-free formulas (see \cite{geneves-rr08} for more details).

\subsection{Compilation of Queries} \label{sec:xpath-compilation}

The logic is expressive enough to capture the set of XPath expressions presented in Section~\ref{sec:queries}.
For example, Figure~\ref{fig:bin} illustrates how the sample XPath 
expression: \begin{center}\texttt{\step{child}{r}[\step{child}{w}/@att]}\end{center} is expressed in the logic. 
From a given context in an XML document, this expression selects all \texttt{r} child nodes which have at least one \texttt{w} child with an attribute \texttt{att}.  Figure~\ref{fig:bin} shows how it is expressed in the logic, on the binary tree representation. The formula holds for \texttt{r} nodes which are 
selected by the expression. The first part of the formula, $\phi$, corresponds to the step \texttt{\step{\axis{child}}{r}} which selects 
candidates \texttt{r} nodes.  The second part, $\psi$, navigates downward in 
the subtrees of these candidate nodes to verify that they have at least one immediate \texttt{w} child with an attribute \texttt{att}.

\begin{figure}[h]
\centering
\begin{small}
\begin{tikzpicture}[scale=1]

\begin{scope}[scale=0.7, xshift=5cm]

\begin{scope}[xshift=-14.25cm, yshift=-2.6cm]
	\draw [\attcolor, dashed]  (14.9,3.6) .. controls +(40:-0.5cm) and +(9:0.9cm) .. (12,4.5); 
	\draw (11.5, 4.55) node (att3) [text= \attcolor] {\labelstyle{att}};
\end{scope}

\draw (2,4) node(x) [text=black] {$\ccontextsymb$};

\draw (3,1) node(c3) [text=purple]{\labelstyle{r}};
\draw (3.4,1) node(c3label) {$_{\phi}$};

\draw (2,2) node(c2) [text=black]{\labelstyle{s}};
\draw [purple] (c2) -- (c3);

\draw (1,3) node(c1)  [text=purple]{\labelstyle{r}};
\draw [purple] (c1) -- (c2);

\draw [purple] (x) -- (c1);

\draw (0,2) node(b1) [text=black]{\labelstyle{v}};
\draw [violet] (c1) -- (b1);

\draw (1,1) node(b2) [text=violet]{\labelstyle{w}};
\draw [violet] (b1) -- (b2);

\draw (1.8,3) node(c1selectedlabel) {$_{\phi \et \psi}$};
\end{scope}

\begin{scope}[scale=1]
\draw (0,-2) node [anchor=north west,  text width=10cm, text badly ragged, at start] (example) {
Translated Query:  \texttt{\textcolor{purple}{\step{child}{\texttt{r}}}\textcolor{violet}{[\step{child}{\texttt{w}}/\textcolor{\attcolor}{\texttt{@att}}]}}  \\~\\Translation:
$\underbrace{\textcolor{purple}{\texttt{r\candop(\clet~\cvar=\cemod{-1}}\textcolor{black}{\ccontextsymb}\texttt{\corop\cemod{-2}\cvar)}}}_{\phi}
\underbrace{\texttt{\textcolor{violet}{\candop\cemod{1}\clet~\cvary=w}\textcolor{\attcolor}{\candop\cemod{att}\ctrue}\textcolor{violet}{\corop\cemod{2}\cvary}}}_{\psi}$};
\end{scope}
\end{tikzpicture}
 \caption{XPath Translation Example.}\label{fig:bin}
 \end{small}
\end{figure}

This example illustrates the need for converse programs inside modalities. The translated XPath expression only uses forward axes (\axis{child} and \axis{attribute}), nevertheless both forward and backward modalities are required for its logical translation.
Without converse programs we would have been unable to differentiate selected nodes from nodes whose existence is simply tested. More generally, properties must often be stated on both the ancestors and the descendants of the selected node. Equipping the logic with both forward and converse programs is therefore crucial. Logics without converse programs may only be used for solving XPath emptiness but cannot be used for solving other decision problems such as containment efficiently.

A systematic translation of XPath expressions into the logic is given in \cite{geneves-pldi07}.  In this paper, we extended it to deal with attributes. We implemented a compiler that takes any expression of the fragment of Figure~\ref{fig:xpath-syntax} and computes its logical translation. With the help of this compiler, we extend the syntax of logical formulas with a logical predicate $\cselectcontext{\exprvar}{\phi}$. This predicate compiles the XPath expression $\exprvar$ given as parameter into the logic, starting from a context that satisfies $\phi$.  The XPath expression to be given as parameter must match the syntax of the XPath fragment shown on Figure~\ref{fig:xpath-syntax} (or Figure~\ref{fig:xpath-sugars}). In a similar manner, we introduce the predicate $\cexistscontext{\exprvar}{\phi}$ which tests the existence of $\exprvar$ from a context satisfying $\phi$, in a qualifier-like manner (without moving to its result). Additionally, the predicate $\cselect{\exprvar}$ is introduced as a shortcut for $\cselectcontext{\exprvar}{\ccontextsymb}$, where $\ccontextsymb$ simply marks the initial context node of the XPath expression\footnote{This mark is especially useful for comparing two or more XPath expressions from the same context.}. The predicate $\cexists{\exprvar}$ is a shortcut for $\cexistscontext{\exprvar}{\ctrue}$. These syntactic extensions of the logic allow the user to easily embed XPath expressions and formulate decision problems out of them (like \eg containment or any other boolean combination). In the next sections we explain how the framework allows combining queries with schema information  for formulating problems.

\subsection{Compilation of Tree Types} \label{sec:xml-schemas-compilation}

Tree type expressions are compiled into the logic in two steps: the first stage translates them into binary tree type expressions, and the second step actually compiles this intermediate representation into the logic. The translation procedure from tree type expressions to binary tree type expressions is well-known and detailed in \cite{geneves-phd06}. The syntax of output expressions follows:
\smallsyntax{
\entry      \treetypevar     [binary tree type expression] 
             \emptyset              [empty set] \oris
        \emptyseq       [empty tree] 
\oris        \treetypevar \tou \treetypevar    [disjunction]
\oris       \treetypelabelwithattributes{\nodelabelvar}{\attributeexprvar}{x, x}      [element definition]
\oris        \typebind{x}{\treetypevar} \treetypevar [binder]
}
Attribute expressions are not concerned by this transformation to binary form: they are simply attached, unchanged, to new (binary) element definitions. 
Finally, binary tree type expressions are compiled into the logic. The logical translation of an expression $\treetypevar$ is given by the function $\translatetype{\treetypevar}{\ctrue}{\cfalse}$ defined below: 
\begin{align*}
\translatetype{\treetypevar}{\phi}{\psi} & \eqdef \cfalse  \quad \text{ for }~~ \treetypevar =\emptyset, \emptyseq\\
\translatetype{\treetypevar_1 \tou \treetypevar_2}{\phi}{\psi}  & \eqdef \translatetype{\treetypevar_1}{\phi}{\psi}  \corop \translatetype{\treetypevar_2}{\phi}{\psi}  \\
\translatetype{ \treetypelabelwithattributes{\nodelabelvar}{\attributeexprvar}{x_1, x_1}}{\phi}{\psi}  & \eqdef  \left( \nodelabelvar  \candop \phi \candop \attrtomu{\attributeexprvar} \candop  \Tsucc{x_1}{1} \candop  \Tsucc{x_2}{2}\right) \corop \psi \\
\translatetype{\typebind{x_i}{\treetypevar_i} \treetypevar}{\phi}{\psi}  & \eqdef  \clet ~ \overline{\cvar_i= \translatetype{\treetypevar_i}{\phi}{\psi} } ~ \cin~ \translatetype{\treetypevar}{\phi}{\psi} 
\end{align*}
where the function $\Tsucc{\cdot}{\cdot}$ sets the type frontier: 
$$\begin{array}{lll}
\Tsucc{x}{\programvar} &=& \left\{ \begin{array}{ll}
\cneg~\cemod{\programvar}\ctrue & \text{\emph{ if }} x \text{\emph{ is bound to }} \emptyseq \\
\cneg~\cemod{\programvar}\ctrue \corop \cemod{\programvar}\cvar & \text{\emph{ if }} \nullable{x} \\ \cemod{\programvar}\cvar & \text{\emph{ if not }} \nullable{x}\end{array}\right.\\
\end{array}$$
according to the predicate $\nullable{x}$ which indicates whether the type $T  \neq \emptyseq$ bound to $x$ contains the empty tree.

The function $\attrtomu{\attributeexprvar}$ compiles attribute expressions associated with element definitions as follows:
\begin{align*}
\attrtomu{\emptyseq} & \eqdef  \negateotheratts{\emptyseq} \\
\attrtomu{\attributeconj \tou \attributeexpr} & \eqdef  \attrtomu{\attributeconj}  \candop \negateotheratts{\attributeconj} \\
\attrtomu{\attributeconj, \attributeconj'} & \eqdef \attrtomu{\attributeconj} \candop \attrtomu{\attributeconj'}\\ 
\attrtomu{\optionalattributevar{\nodelabelvar}} & \eqdef \nodelabelvar \corop \cneg~\nodelabelvar \\
\attrtomu{\attributevar{\nodelabelvar} } & \eqdef \nodelabelvar \\
\attrtomu{\neg \attributevar{\nodelabelvar}} & \eqdef \cneg~\nodelabelvar
\end{align*}
In usual schemas (\eg DTDs, XML Schemas) when no attribute is specified for a given element, it simply means no attribute is allowed for the defined element. This convention must be explicitly stated into the logic. This is the role of the function ``$\negateotheratts{\attributeconj}$'' which returns the negated disjunction of all attributes not present in $\attributeconj$. As a result, taking attributes into account comes at an extra-cost. The above translation appends a (potentially very large) formula in which all attributes occur,  
for each element definition. In practice, 
a placeholder atomic proposition is inserted until the full set of attributes involved in the problem formulation is known. When the whole formula has been parsed, 
placeholders are replaced by the conjunction of negated attributes they denote.  This extra-cost can be observed in practice, and the system allows two modes of operations: with or without attributes\footnote{The optional argument ``-attributes'' must be supplied for attributes to be considered.}. Nevertheless the system is still capable of handling real world DTDs (such as the DTD of XHTML 1.0 Strict) with attributes. This is due to (1) the limited expressive power of languages such as DTD that do not allow for disjunction over attribute expressions (like ``$\attributeconj \tou \attributeexpr$'' ); and, more importantly, (2) the satisfiability-testing algorithm which is implemented using symbolic techniques \cite{geneves-rr08}.

Tree type expressions form the common internal representation for a variety of XML schema definition languages. In practice, the logical translation of a tree type expression $\treetypevar$ are obtained directly from a variety of formalisms for defining schemas, including DTD, XML Schema, and Relax NG. For this purpose, the syntax of logical formulas is extended with a predicate  $\ctype{\cdot}{\cdot}$. The logical translation of an existing schema is returned by $\ctype{\filenamevar}{\nodelabelvar}$ where $\filenamevar$ is a file path to the schema file and $\nodelabelvar$ is the element name to be considered as the entry point (root) of the given schema. Any occurence of this predicate will parse the given schema, extract its internal tree type representation $\treetypevar$, compile it into the logic and return the logical formula $\translatetype{\treetypevar}{\ctrue}{\cfalse}$.

\subsection{Type Tagging}

A tag (or ``color'') is introduced in the compilation of schemas with the purpose of marking all node types of a specific schema. A tag is simply a fresh atomic proposition passed as a parameter to the translation of a tree type expression. For example: 
$\translatetype{\treetypevar}{\text{xhtml}}{\cfalse}$ is the logical translation of $\treetypevar$ where each element definition is annotated with the atomic proposition ``\text{xhtml}''.  With the help of tags, it becomes possible to refer to the element types in any context. 
For instance, one may formulate
$\translatetype{\treetypevar}{\text{xhtml}}{\cfalse} \corop \translatetype{\treetypevar'}{\text{smil}}{\cfalse}$ for denoting the union of all $\treetypevar$ and $\treetypevar'$ documents, while keeping a way to distinguish element types; even if some element names are shared by the two type expressions. 

Tagging becomes even more useful for characterizing evolutions between successive versions of a single schema. In this setting, we need a way to distinguish nodes allowed by a newer schema version from nodes allowed by an older version. This distinction must not be based only on element names, but also on content models. Assume for instance that $\treetypevar'$ is a newer version of schema $\treetypevar$. If we are interested in the set of trees allowed by $\treetypevar'$ but not allowed by $\treetypevar$ then we may formulate:
$$\translatetype{\treetypevar'}{\ctrue}{\cfalse} \candop \cneg~\translatetype{\treetypevar}{\ctrue}{\cfalse}$$
If we now want to check more fine-grained properties, we may rather be interested in the following (tagged) formulation:
$$\translatetype{\treetypevar'}{\text{all}}{\cfalse} \candop \cneg~\translatetype{\treetypevar}{\ctrue}{\cneg~\text{old\_complement}}$$
In this manner, we can distinguish elements that were added in $\treetypevar'$ and whose names did not occur in $\treetypevar$, from elements whose names already occured in  $\treetypevar$ but whose content model changed in $\treetypevar'$, for instance.  
In practice, a type is tagged using the predicate $\ctypetag{\filenamevar}{\nodelabelvar}{\phi}{\phi'}$ which parses the specified schema, converts it into its logical representation $\treetypevar$ and returns the formula $\translatetype{\treetypevar}{\phi}{\phi'}$. Such kind of type tagging is useful for studying the consequences of schema updates over queries, as presented in the next sections.

\section{Analysis Predicates} \label{sec:predicates}

This section introduces the basic analysis tasks offered to XML application designers for
assessing the impact of schema evolutions. In particular, we propose a mean for identifying
the precise reasons for type mismatches or changes in query 
results under type constraints. 

For this purpose, we build on our query and type expression compilers, and define additional predicates that facilitate the formulation of decision problems at a higher level of abstraction. Specifically, these predicates are introduced as logical macros with the goal of allowing system usage while focusing (only) on the XML-side properties, and keeping underlying logical issues transparent for the user. Ultimately, we regard the set of basic logical formulas (such as modalities and recursive binders) as an assembly language, to which  predicates are translated.

We illustrate this principle with two simple predicates designed for checking backward-compatibility of schemas, and query satisfiability in the presence of a schema.

\begin{itemize}
\item The predicate $\cbackwardincompatible{\treetypevar}{\treetypevar'}$ takes two type expressions as parameters, and assumes $\treetypevar'$ is  an altered version of $\treetypevar$. This predicate is unsatisfiable iff all instances of $\treetypevar'$ are also valid against $\treetypevar$. Any occurrence of this predicate in the input formula will automatically be compiled as $\translatetype{\treetypevar'}{\ctrue}{\cfalse} \candop \cneg~\translatetype{\treetypevar}{\ctrue}{\cfalse}$.

\item The predicate $\cnonempty{\exprvar}{\treetypevar}$ takes an XPath expression (with the syntax defined on Figure~\ref{fig:xpath-syntax}) and a type expression as parameters, and is unsatisfiable iff the query always returns an empty set of nodes when evaluated on an XML document valid against $\treetypevar$. This predicate compiles into 
$\cselectcontext{\exprvar}{\translatetype{\treetypevar}{\ctrue}{\cfalse} \candop \ccontextsymb}$
where the predicate $\cselectcontext{\exprvar}{\phi}$ compiles the XPath expression $\exprvar$ into the logic, starting from a context that satisfies $\phi$, as explained in Section~\ref{sec:xpath-compilation}. This can be used to check whether the modification of the schema does not contradict any part of the query.
\end{itemize}

Notice that the predicate $\cnonempty{\exprvar}{\treetypevar}$ can be used for checking whether a query that is valid\footnote{We say that a query is \emph{valid} iff its negation is unsatisfiable.} against a schema remains valid with an updated version of a schema. In other terms, this predicate allows determining whether a query that must always return a non-empty result (whatever the tree on which it is evaluated) keeps verifying the same property with a new version of a schema.

A second, more-elaborated, class of predicates allows formulating problems that combine both a query $\exprvar$ and two type expressions $\treetypevar, \treetypevar'$ (where $\treetypevar '$ is assumed to be a evolved version of $\treetypevar$):

\begin{itemize}
\item $\cnewelementnames{\exprvar}{\treetypevar}{\treetypevar'}$ is satisfied iff the query $\exprvar$ selects elements whose names did not occur at all in $\treetypevar$. This is especially useful for queries whose last navigation step contains a ``\texttt{*}'' node test and may thus select unexpected elements. This predicate is compiled into: $$\cneg \celem{\treetypevar}{} \candop \cselectcontext{\exprvar}{ \translatetype{\treetypevar'}{\ctrue}{\cfalse}}$$
where $\celem{\treetypevar}$ is another predicate that builds the disjunction of all element names occuring in $\treetypevar$. In a similar manner, the predicate $\cattr{\phi}$ builds the logical disjunction of all attribute names used in $\phi$.
%
%
\item $\cnewregions{\exprvar}{\treetypevar}{\treetypevar'}$ is satisfied iff the query $\exprvar$ selects elements whose names already occurred  in $\treetypevar$, but such that these nodes now occur in a new context in $\treetypevar'$. In this setting, the path from the root of the document to a node selected by the XPath expression $\exprvar$ contains a node whose type is defined in $\treetypevar'$ but not in $\treetypevar$ as illustrated below:
\begin{center}
\begin{tikzpicture}[scale=0.75]

\begin{scope}[scale=1]
\draw [black] (3.25,0) -- (3.75,0) -- (6, -4) -- (1, -4) --cycle;

\draw (3.1,-3) node (a) {};
\draw (a) -- (5,-4)-- (2,-4) --cycle;
\fill (3.22,-3.02) circle (0.09cm);

\draw [\changingcolor, dashed, thick]  (3.5,0) .. controls +(40:-0.5cm) and +(9:0.9cm) .. (a);

\draw (0,-2) node[anchor=north west,  text width=2cm, text badly ragged, at start]  (comment1) {node selected by $\exprvar$};
\draw[->] (comment1) -- (a);

\begin{scope}[xshift=5.5cm, yshift=-0.5cm]
\draw (0,0) node [anchor=north west,  text width=2.25cm, text badly ragged, at start]  (comment2) {path from root to selected node contains  node in $\treetypevar' \setminus \treetypevar$};
\draw[->] (comment2) -- (-1.5, -1.5);
\end{scope}

\begin{scope}[xshift=1.25cm, yshift=-4cm]
\draw (0,0) node [anchor=north west,  text width=5cm, text badly ragged, at start] (comment3) {XML document valid against $\treetypevar'$ but not against $\treetypevar$};
\end{scope}

\end{scope}
\end{tikzpicture}
\end{center}

The predicate $\cnewregions{\exprvar}{\treetypevar}{\treetypevar'}$ is logically defined as follows: 
\begin{multline*}
\cnewregions{\exprvar}{\treetypevar}{\treetypevar'} 
\eqdef \\
\cselectcontext{\exprvar}{\translatetype{\treetypevar}{\call}{\cfalse} \candop \cneg~ \translatetype{\treetypevar'}{\ctrue}{\cneg~ \coldcomplement}} \\
\candop \cneg~ \caddedelement{\treetypevar}{\treetypevar'} \\
\candop \cancestor{\coldcomplement} \\
\candop \cneg~ \cdescendant{\coldcomplement} \\
\candop \cneg~ \cfollowing{\coldcomplement} \\
\candop \cneg~ \cpreceding{\coldcomplement} \\
\end{multline*}
The previous definition heavily relies on the partition of tree nodes defined by XPath axes, as illustrated by Figure~\ref{fig:xpath-axes-partition}. The definition of  $\cnewregions{\exprvar}{\treetypevar}{\treetypevar'}$ uses an auxiliary predicate $\caddedelement{\treetypevar}{\treetypevar'}$ that builds the disjunction of all element names defined in $\treetypevar'$ but not in $\treetypevar$ (or in other terms, elements that were added in $\treetypevar'$). 
In a similar manner, the predicate $\caddedattribute{\phi}{\phi'}$ builds the disjunction of all attribute names defined in $\treetypevar'$ but not in $\treetypevar$.

\begin{figure}[h]
\begin{center}
\includegraphics[keepaspectratio=true, width=8cm]{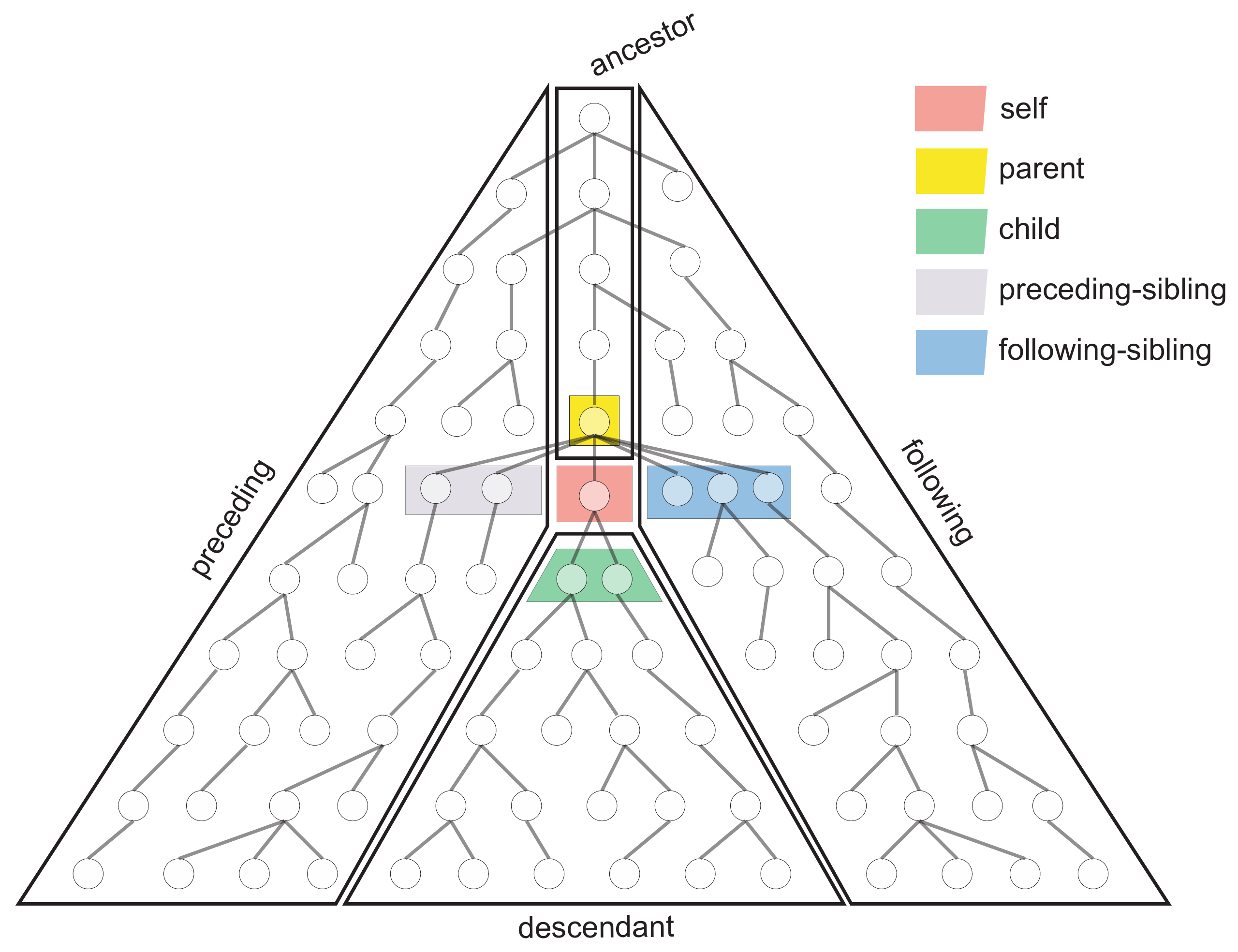}
\end{center}
\caption{XPath axes: partition of tree nodes.}\label{fig:xpath-axes-partition}
\end{figure}

The predicate $\cnewregions{\exprvar}{\treetypevar}{\treetypevar'}$ is useful for checking whether a query selects a different set of nodes with $\treetypevar'$ than with $\treetypevar$ because selected elements  may occur in new regions of the document due to changes brought by $\treetypevar'$.

\item $\cnewcontents{\exprvar}{ \treetypevar}{\treetypevar '}$ is satisfied iff the query $\exprvar$ selects elements whose names were already defined in $\treetypevar$, but whose content model has changed due to evolutions brought by $\treetypevar'$, as illustrated below:
\begin{center}
\begin{tikzpicture}[scale=0.75]

\begin{scope}[scale=1]
\draw [black] (3.25,0) -- (3.75,0) -- (6, -4) -- (1, -4) --cycle;

\draw (3.1,-3) node (a) {};
\draw [fill=\changingcolor] (a) -- (5,-4)-- (2,-4) --cycle;
\fill (3.22,-3.02) circle (0.09cm);

\draw [black, dashed]  (3.5,0) .. controls +(40:-0.5cm) and +(9:0.9cm) .. (a);

\draw (0,-2) node[anchor=north west,  text width=2cm, text badly ragged, at start]  (comment1) {node selected by $\exprvar$};
\draw[->] (comment1) -- (a);

\begin{scope}[xshift=5.5cm, yshift=-0.5cm]
\draw (0,0) node [anchor=north west,  text width=2.25cm, text badly ragged, at start]  (comment2) {subtree for selected node has changed (new content model)};
\draw[->] (comment2) -- (-1.15, -2.85);
\end{scope}

\begin{scope}[xshift=1.25cm, yshift=-4cm]
\draw (0,0) node [anchor=north west,  text width=5cm, text badly ragged, at start] (comment3) {XML document valid against $\treetypevar'$ but not against $\treetypevar$};
\end{scope}
\end{scope}
\end{tikzpicture}
\end{center}
The definition of $\cnewcontents{\exprvar}{ \treetypevar}{\treetypevar '}$ follows: 
\begin{multline*}
\cnewcontents{\exprvar}{\treetypevar}{\treetypevar'} 
\eqdef \\
\cselectcontext{\exprvar}{\translatetype{\treetypevar}{\call}{\cfalse} \candop \cneg~ \translatetype{\treetypevar'}{\ctrue}{\cneg~ \coldcomplement}} \\
\candop \cneg~ \caddedelement{\treetypevar}{\treetypevar'} \\
\candop \cneg~ \cancestor{ \caddedelement{\treetypevar}{\treetypevar'}} \\
\candop \cdescendant{\coldcomplement} \\
\candop \cneg~ \cfollowing{\coldcomplement} \\
\candop \cneg~ \cpreceding{\coldcomplement} \\
\end{multline*}
The predicate $\cnewcontents{\exprvar}{ \treetypevar}{\treetypevar '}$ can be used for ensuring that XPath expressions will not return nodes with a possibly new content model that may cause problems. For instance, this allows checking whether an XPath expression whose resulting node set is converted to a string value (as in, \eg XPath expressions used in XSLT ``value-of'' instructions) is affected by the changes from $\treetypevar $ to $\treetypevar'$.
%
\end{itemize}
The previously defined predicates can be used to help the programmer identify precisely how type constraint evolutions affect queries. They can even be combined with usual logical connectives to formulate even more sophisticated problems.  
%
%
%
For example, let us define the predicate $\cexclude{\phi}$ which is satisfiable iff there is no node that satisfies $\phi$ in the whole tree. This predicate can be used for excluding specific element names or even nodes selected by a given XPath expression. It is defined as follows:
\begin{multline*}
\cexclude{\phi} \eqdef \\ \cneg~ \cancestororself{\cdescendantorself{\phi}}
\end{multline*}
This predicate can also be used for checking properties in an iterative manner, refining the property to be tested at each step. It can also be used for verifying fine-grained properties. For instance, one may check whether $\treetypevar'$ defines the same set of trees as $\treetypevar$ modulo new element names that were added in $\treetypevar'$ with the following formulation:
$$\cneg~(\treetypevar \cequiv \treetypevar') \candop \cexclude{\caddedelement{\treetypevar}{\treetypevar'}}$$ This allows identifying that, during the type evolution from $\treetypevar$ to $\treetypevar'$, the query results change has not been caused by the type extension but by new compositions of nodes from the older type.

In practice, instead of taking internal tree type representations (as defined in Section~\ref{sec:tree-type-expressions}) as parameters,  most predicates do actually take any logical formula as parameter,  or even schema paths as parameters. We believe this facilitates predicates usage and, most notably, how they can be composed together. Figure~\ref{fig:predicates-syntax} gives the syntax of built-in predicates as they are implemented in the system, where $\filenamevar$ is a file path to a DTD (.dtd), XML Schema (.xsd), or Relax NG (.rng).
\begin{figure}[h]
\begin{center}
\smallsyntax{ 
\entry \cpredicatevar [] 
\cselect{\exprvar} [] 
\oris        \cselectcontext{\exprvar}{\phi} []
\oris        \cexists{\exprvar} []
\oris        \cexistscontext{\exprvar}{\phi} [] \\
\oris       \ctype{\filenamevar}{\nodelabelvar} [] 
\oris       \ctypetag{\filenamevar}{\nodelabelvar}{\phi}{\phi'} [] 
\oris       \cforwardincompatible{\phi}{\phi'}   []
\oris       \cbackwardincompatible{\phi}{\phi'}   []\\
\oris       \celem{\phi} []
\oris       \cattr{\phi} [] 
\oris       \cdescendant{\phi} []
\oris       \cexclude{\phi} []
\oris       \caddedelement{\phi}{\phi'} []
\oris       \caddedattribute{\phi}{\phi'} [] \\
\oris       \cnonempty{\exprvar}{\phi} []
\oris       \ccnewelementnames{\exprvar}{\filenamevar}{\filenamevar'}{\nodelabelvar}  []
\oris       \ccnewregions{\exprvar}{\filenamevar}{\filenamevar'}{\nodelabelvar}   []
\oris       \ccnewcontents{\exprvar}{\filenamevar}{\filenamevar'}{\nodelabelvar}  []
\oris \custompredicatevar(\oneormoresep{\phi}) []
}
\end{center}
\caption{Syntax of Predicates for XML Reasoning.}\label{fig:predicates-syntax}
\end{figure}
In addition of aforementioned predicates, the predicate $\cdescendant{\phi}$ forces the existence of a node satisfying $\phi$ in the subtree, and $\custompredicatevar(\oneormoresep{\phi})$ is a call to a custom predicate, as explained in the next section.

\subsection{Custom Predicates}

Following the spirit of predicates presented in the previous section, users may also define their own custom predicates.  The full syntax of XML logical specifications to be used with the system is defined on Figure~\ref{fig:lang-syntax}, where the meta-syntax $\oneormoresep{X}$ means one or more occurrence of $X$ separated by commas.
A global problem specification can be any formula (as defined on Figure~\ref{fig:formulas-syntax}), or a list of custom predicate definitions separated by semicolons and followed by a formula.
A custom predicate may have parameters that are instanciated with actual formulas when the custom predicate is called (as shown on Figure~\ref{fig:predicates-syntax}). A formula bound to a custom predicate may include calls to other predicates, but not to the currently defined predicate (recursive definitions must be made through the let binder shown on Figure~\ref{fig:formulas-syntax}). 

\begin{figure}[h]
\begin{center}
\smallsyntax{
\entry \logicalspec []
 \phi [ formula (see Fig.~\ref{fig:formulas-syntax})]
\oris \predicatedefinitions; \phi []\\
\entry \predicatedefinitions []
 \custompredicatevar(\oneormoresep{\nodelabelvar}) = \phi'  [custom definition]
\oris \predicatedefinitions; \predicatedefinitions [list of definitions]
}
\end{center}
\caption{Global Syntax for Specifying Problems.}\label{fig:lang-syntax}
\end{figure}

\section{Framework in Action} \label{sec:applications-xml-evolution}

We have implemented the whole software architecture described in Section~\ref{sec:framework} and illustrated on Figure~\ref{fig:framework} \cite{implementation}. We have carried out extensive experiments of the system with real world schemas such as XHTML, MathML, SVG, SMIL (Table~\ref{table:sizes} gives details related to their respective sizes) and queries found in transformations such MathML content to presentation \cite{mathmlc2p}. We present two of them that show how the tool can be used to analyze different situations where schemas and queries evolve.  

\begin{table}
\begin{tabular}{lccc}
Schema & Variables & Elements & Attributes \\ \hline
XHTML 1.0 basic DTD &  71  & 52 & 57 \\
XHTML 1.1 basic DTD &  89  & 67 & 83 \\
MathML 1.01 DTD & 137 &127 & 72 \\
MathML 2.0 DTD & 194 &181 & 97
\end{tabular}
\caption{Sizes of (Some) Considered Schemas.}\label{table:sizes}
\end{table}

\subsection*{Evolution of XHTML Basic}

The first test consists in analyzing the relationship (forward and backward compatibility) between XHTML basic 1.0 and XHTML basic 1.1 schemas. In particular, backward compatibility can be checked by the following command:
\begin{verbatim}
backward_incompatible("xhtml-basic10.dtd", 
                      "xhtml-basic11.dtd", "html") 
\end{verbatim}

The test immediately yields a counter example as the new schema contains new element names. The counter example (shown below) contains a \texttt{style}  element occurring as a child of \texttt{head}, which is not permitted in XHTML basic 1.0:
\begin{Verbatim}[frame=single, fontsize=\scriptsize]
<html>
  <head>
    <title/>
    <style type="_otherV"/>
  </head>
  <body/>
</html>
\end{Verbatim}

The next step consists in focusing on the relationship between both schemas excluding these new elements. This can be formulated by the following command:
\begin{verbatim}
backward_incompatible("xhtml-basic10.dtd",
                      "xhtml-basic11.dtd", "html")
& exclude(added_element(
      type("xhtml-basic10.dtd","html"), 
      type("xhtml-basic11.dtd", "html")))
\end{verbatim}
The result of the test  shows a counter example document that proves that XHTML basic 1.1 is not backward compatible with XHTML basic 1.0 even if new elements are not considered. In particular, the content model of the \texttt{label} element cannot have an \texttt{a} element in XHTML basic 1.0
while it can in XHTML basic 1.1. The counter example produced by the solver is shown below:
\begin{Verbatim}[frame=single, fontsize=\scriptsize]
<html>
  <head>
    <object>
      <label>
        <a>
          <img/>
        </a>
        <img/>
      </label>
      <param/>
    </object>
    <meta/>
    <title/>
    <base/>
  </head>
  <body/>
</html>

XTML basic 1.0 validity error: element "a" is not declared
in "label" list of possible children
\end{Verbatim}
Notice that we observed similar forward and backward compatibility issues with several other W3C normative schemas (in particular for the different versions of SMIL and SVG). Such backward incompatibilities suggests that applications cannot simply ignore new elements from newer schemas, as the combination of older elements may evolve significantly from one version to another.

\subsection*{MathML Content to Presentation Conversion}

MathML is an XML format for describing mathematical notations and capturing both its 
structure and graphical structure, also known as Content MathML and Presentation MathML respectively.
The structure of a given equation is kept separate from the presentation and the rendering part can be generated
from the structure description. This operation is usually carried out using an XSLT transformation
that achieves the conversion. In this test series, we focus on the analysis
of the queries contained in such a transformation sheet and evaluate the impact of the schema change from 
MathML 1.0 to MathML 2.0 on these queries.

Most of the queries contained in the transformation represent only a few patterns very similar up to element 
names. The following three  patterns are the most frequently used:
\begin{verbatim}
Q1:  //apply[*[1][self::eq]]
Q2:  //apply[*[1][self::apply]/inverse]
Q3:  //sin[preceding-sibling::*[position()=last()
             and (self::compose or self::inverse)]]
\end{verbatim}
The first test is formulated by the following command:
\begin{verbatim}
new_region("Q1","mathml.dtd","mathml2.dtd","math")
\end{verbatim}
The result of the test  shows a counter example document that proves that the query may select nodes in new contexts in MathML 2.0 compared to MathML 1.0. In particular, the  query \texttt{Q1}  selects \texttt{apply}
elements whose ancestors can be \texttt{declare} elements, as indicated on the document produced by the solver:
\begin{Verbatim}[frame=single, fontsize=\scriptsize]
<math xmlns:solver="http://wam.inrialpes.fr/xml"
            solver:context="true">
  <declare>
    <apply solver:target="true">
      <eq/>
    </apply>
    <condition/>
  </declare>
</math>
\end{Verbatim}
Notice that the solver automatically annotates a pair of nodes related by the query: when the query is evaluated from a node marked with the attribute \texttt{solver:context}, the node marked with \texttt{solver:target} is selected. To evaluate the effect of this change, the counter example is filled with content and passed as an input parameter to the transformation. This shows
immediately a bug in the transformation as the resulting document is not a MathML 2.0 presentation document. Based on this analysis, we know
that the XSLT template associated with the match pattern \texttt{Q1} must be updated to cope with MathML evolution from version 1.0 to version 2.0.

The next test consists in evaluating the impact of the MathML type evolution for the query \texttt{Q2} while excluding all new elements added in MathML 2.0 from the test.
This identifies whether old elements of MathML 1.0 can be composed in MathML 2.0 in a different manner. This can be performed with the following command:
\begin{verbatim}
new_content("Q2","mathml.dtd","mathml2.dtd","math") 
& exclude(added_element(type("mathml.dtd","math"),
                        type("mathml2.dtd", "math")))
\end{verbatim}

The test result shows an example document that effectively combines MathML 1.0 elements in a way
that was not allowed in MathML 1.0 but permitted in MathML 2.0.
\begin{Verbatim}[frame=single, fontsize=\scriptsize]
<math xmlns:solver="http://wam.inrialpes.fr/xml"
            solver:context="true">
  <apply solver:target="true">
    <apply>
      <inverse/>
    </apply>
    <annotation-xml>
      <math/>
    </annotation-xml>
    <condition/>
  </apply>
</math>
\end{Verbatim}

Similarly, the last test consists in evaluating the impact of the MathML type evolution for the query \texttt{Q3},  excluding all new elements added in MathML 2.0 and counter example documents containing \texttt{declare} elements (to avoid trivial counter examples):
\begin{verbatim}
new_regions("Q3","mathml.dtd","mathml2.dtd","math")
& exclude(added_element(type("mathml.dtd","math"),
                        type("mathml2.dtd","math"))) 
                        & exclude(declare) 
\end{verbatim}
The counter example document shown below illustrates a case where the \texttt{sin} element occurs in a new context.
\begin{Verbatim}[frame=single, fontsize=\scriptsize]
<math xmlns:solver="http://wam.inrialpes.fr/xml"
            solver:context="true">
  <apply>
    <annotation-xml>
      <math>
        <apply>
          <inverse/>
          <sin solver:target="true"/>
        </apply>
      </math>
    </annotation-xml>
  </apply>
</math>
\end{Verbatim}

Applying the transformation on previous examples yields documents which are neither MathML 1.0 nor MathML 2.0 valid.
As a result, the stylesheet cannot be used safely over documents of the new type without modifications. In addition, the required changes 
to the stylesheet are not limited to the addition of new templates for MathML 2.0 elements. The templates that deal with 
the composition of MathML 1.0 elements should be revised as well.

All the previous tests were processed in less than 30 seconds on an ordinary laptop computer running Java under Mac OS X.

\section{Conclusion} \label{conclusion}

In this article, we present a logical framework and a tool for verifying forward/backward compatibility issues
caused by schemas and queries evolution. The tool allows XML designers to identify 
queries that must be reformulated in order to produce the expected results across successive schema versions. 
With this tool designers can examine precisely the impact of schema changes over queries, therefore facilitating their 
reformulation. We gave illustrations of how to use the tool for both schema and query evolution on realistic examples. 
In particular,  we considered typical situations in applications involving 
W3C schemas evolution  such as  XHTML and MathML. The tool can be very useful  
for standard schema writers and maintainers in order to assist them enforce some level of quality assurance on compatibility between versions.

There are a number of interesting extensions to the proposed system. First, the set of predicates can be easily 
enriched to detect more precisely the impact on queries. For example, one can extend the tagging to identify separately 
every navigation step  and qualifier in a query expression. This will help greatly in the identification and reformulation of 
the navigation steps or qualifiers affected by schemas evolution.

\bibliographystyle{abbrv}
\bibliography{references} 

\end{document}